\documentclass[twocolumn,nofootinbib,superscriptaddress]{revtex4-1}
\usepackage[T2A]{fontenc}
\usepackage[utf8]{inputenc}
\usepackage{graphicx}
\usepackage[colorlinks,citecolor=blue,urlcolor=blue,linkcolor=blue]{hyperref}
\usepackage[utf8]{inputenc}
\usepackage{csquotes}
\usepackage{amsfonts,amsmath,amssymb,mathrsfs}
\usepackage{tempora}        
\usepackage{mathptmx}       
\usepackage{physics}
\newcommand{\kap}{\varkappa}
\newcommand{\eps}{\varepsilon}
\newcommand{\wt}[1]{{\widetilde #1}}
\newcommand{\ovl}[1]{{\overline #1}}
\newcommand{\Ol}[1]{\mkern3.5mu\overline{\mkern-3.5mu #1 \mkern-0.5mu}\mkern0.5mu}
\DeclareMathOperator{\sgn}{sgn}
\begin{document}
\title{\bf On the resonances near the continua boundaries of the Dirac equation with a short-range interaction}
\author{K.~S.~Krylov}
\email[E-mail: ]{krylov@theor.mephi.ru}
\affiliation{National Research Nuclear University MEPhI, Moscow 115409, Russia}
\author{V.~D.~Mur}
\affiliation{National Research Nuclear University MEPhI, Moscow 115409, Russia}
\author{A.~M.~Fedotov}
\affiliation{National Research Nuclear University MEPhI, Moscow 115409, Russia}
\date{\today}
\begin{abstract}
Using the model of a deep spherically symmetric rectangular well as an example, it is shown that resonant scattering near the boundaries of the lower or upper Dirac continua cannot serve as evidence in favor of spontaneous electron-positron pair production in a supercritical domain.
\end{abstract}
\maketitle
\section{Introduction}
\label{intro}
Apparently, the applicability of a one-particle approach to relativistic wave equations in a strong electrostatic field decaying at infinity was first questioned and discussed by Schiff, Snyder and Weinberg \cite{SchiffSnyderWeinberg1940PhysRev}. Using an example of the Klein--Fock--Gordon (KFG) equation in a spherically symmetric rectangular potential well of depth $V$, they demonstrated that complex energy eigenvalues emerge once $V$ exceeds a critical value. This means that for such values of $V$ the Hamiltonian of the Pauli--Weisskopf theory cannot be diagonalized, hence is not a self-adjoint operator.

This statement was extended to the case of an arbitrary potential shape by employing the effective radius approximation \cite{LandauLifshitz3_1981Butterworth_eng, Newton1966McGrawHill_eng}, generalized in \cite{MurPopov1976TheorMathPhys_BosonCase} to the case of a relativistic scalar field. As shown in \cite{MurPopov1976TheorMathPhys_BosonCase}, at a certain value of $V = V_c > V_0$\footnote{The value $V = V_0$ is such that a bound state emerges at the boundary of the lower continuum. The levels emerging from there should be interpreted as bound states of antiparticles \cite{Migdal1972JETP_1973NuclPhysB_eng}.}, the two branches of the one-particle spectrum (for particles and antiparticles) do overlap. The formal continuation of one-particle solutions to the region $V > V_c$ leads to the states with complex energy on a physical sheet. This indicates that the Hamiltonian of the KFG equation (without taking into account vacuum polarization) for $V > V_c$ ceases to be self-adjoint, see also \cite{SchroerSwieca1970PhysRevD}.

Unlike scalar particles, electronic levels\footnote{In \cite{SchiffSnyderWeinberg1940PhysRev}, the levels with Dirac quantum number $\kap = \pm 1$ were discussed.} as $V$ increases move from the bottom of the upper continuum of solutions of the Dirac equation to the upper boundary $E = -mc^2$ of the lower one. However, the authors of \cite{SchiffSnyderWeinberg1940PhysRev} believed that for sufficiently large $V$ the vacuum is no longer the lowest energy state, but differs from that by finite amounts of charge and energy.

In more detail (for any values of $\kap$) this problem is considered in the monograph \cite{AkhiezerBerestetskii1981QED}, where both cases of wide and narrow well (as compared to the electron Compton length $l_C = \hbar/mc$) are discussed. It is still stated there that a difficulty in interpretation arises when $V$ exceeds the so-called critical value $V_c$, at which a given level reaches the lower continuum boundary. By analogy with the works \cite{PomeranchukSmorodinsky1945JPhysUSSR, ZeldovichPopov1972SovPhysUsp}, it is argued that this difficulty is associated with electron-positron pair production by the field at $V > V_c$, so that the problem \enquote{cannot be solved within a scope of quantum mechanics of a single particle}.

However, such an interpretation of the phenomena with $V > V_c$ is illegal \cite{MurPopov1976TheorMathPhys_FermionCase}. A crucial difference between the properties of solutions of the KFG and Dirac equations can be seen by considering motion of the poles of the scattering matrix near the boundary of the lower continuum as the depth of the well is varied\footnote{In \cite{MurPopov1976TheorMathPhys_FermionCase} this was done for the Dirac equation with a short-range potential using the effective radius approximation.}. While in scalar case, after a collision at $V = V_c$, upon $V > V_c$ a pair of poles diverges in the upper half-plane of complex wavevector $k$, in spinor case this difficulty never occurs, since after the collision the poles of the $S$-matrix depart to the lower half-plane of the complex $k$-plane, i.e. to the second, unphysical sheet, cf. Figs.~1 and 2 in \cite{MurPopov1976TheorMathPhys_FermionCase}. This is due to the fact that, by virtue of the Pauli exclusion principle for fermions, one can still stay within a single-particle approach by defining a \enquote{Dirac sea} distorted by an external field.

A similar situation arises in a relativistic Coulomb problem, first considered by Pomeranchuk and Smorodinsky \cite{PomeranchukSmorodinsky1945JPhysUSSR}\footnote{It was this work where the term \enquote{critical charge} was introduced.}, for which we conventionally denote nuclear charge by $Z$. It is well known \cite{AkhiezerBerestetskii1981QED} that a model of pointlike Coulomb potential is an excessive idealization loosing sense at the singular value $Z_s = \alpha^{-1} |\kap| \approx 137 |\kap|$, where $\alpha = e^2/\hbar c$ is the Sommerfeld fine structure constant. But regularization of the Coulomb potential at small distances by accounting for a finite nucleus size removes this difficulty. The value $Z_\text{cr}$, for which a ground level reaches the boundary of the lower continuum before disappearing from the spectrum, was initially calculated in \cite{PomeranchukSmorodinsky1945JPhysUSSR} using the simplest \enquote{rectangular cutoff} but was later refined to $Z_\text{cr} \approx 170$ in \cite{PieperGreiner1969ZPhys, Rein1969ZPhys, Popov1970JETPLett}. Note that motion of the levels in the Coulomb problem with $Z \leqslant Z_\text{cr}$ is quite the same as in a rectangular well with $V \leqslant V_c$ considered originally in \cite{SchiffSnyderWeinberg1940PhysRev}.

Later on, these issues were extensively studied by numerous authors, see the review \cite{ZeldovichPopov1972SovPhysUsp}, the monograph \cite{GreinerMullerRafelski1985} and references therein. According to all these works, a \enquote{bare} nucleus with $Z > Z_\text{cr}$ spontaneously produces electron-positron pairs, with positrons departing to infinity, and the electrons shielding the nuclear charge, which is equivalent to acquiring a negative charge by the vacuum. However, the work \cite{KuleshovMurEtAl2015PhysicsUspekhi} demonstrated that the phenomena occurring when the nuclear charge $Z > Z_\text{cr}$ must be interpreted differently.

Let us recall its main argument in a greater length. A discrete level, after reaching the boundary of the lower continuum at $Z = Z_\text{cr}$, collides with a virtual one and disappears from the spectrum. These levels, corresponding to the poles of the scattering matrix $S_\kap = \exp[2i\delta_\kap (k)]$ and initially moving towards each other along the imaginary axis of the $k$-plane, after a collision upon $Z > Z_\text{cr}$ go into the lower half-plane of the complex $k$-plane, i.e. to the unphysical sheet. Their continuation is a pair of Breit--Wigner poles. A pole closest to the physical domain with energy $E_\text{BW} = -E_0 + i\Gamma/2$ ($E_0 > 0$ and $\Gamma > 0$) corresponds to a quasistationary state in the lower continuum with energy $\Ol{E}_\text{BW} = E_0 - i\Gamma/2$, manifesting itself as a Breit--Wigner resonance in scattering of positrons with energy $\Ol{E} = -E$. Then $E_0$ is the position and $\Gamma$ the width of the quasidiscrete level. In the aforementioned works, the width of such a Breit--Wigner resonance was interpreted as the probability of spontaneous production of $e^+e^-$-pairs. But such interpretation contradicts the general principles of quantum theory. Indeed, the radial Dirac Hamiltonian with Coulomb potential regularized at small distances is a self-adjoint operator \cite{KuleshovMurEtAl2015PhysicsUspekhi}, see \cite{VoronovGitmanTyutin2007TheorMathPhys, KuleshovMurEtAl2017JETP} for details. Therefore, quantization based on a full set of its solutions, i.e. the Furry picture \cite{Furry1951PhysRev}, is legal. Herewith the scattering states in the lower continuum (i.e. in the Dirac sea of the non-second quantized approach) correspond to positron scattering. The fact that the partial phase $\delta_\kap(k)$ is real ensures unitarity of the partial matrix $S_\kap(k)$ of elastic scattering, hence absence of any inelastic processes including spontaneous production of $e^+e^-$-pairs. Note that all the formal results of \cite{KuleshovMurEtAl2015PhysicsUspekhi} have been confirmed in \cite{GodunovMachetVysotsky2017EurPhysJ} by independent calculations.

However, the authors of \cite{GodunovMachetVysotsky2017EurPhysJ} did not share the conclusion of \cite{KuleshovMurEtAl2015PhysicsUspekhi}, still retaining a belief that resonances in scattering of positrons by a supercritical nucleus are namely the indicators of spontaneous production of $e^+e^-$-pairs and leaning towards the description of the phenomena with $Z > Z_\text{cr}$ stated in the review \cite{ZeldovichPopov1972SovPhysUsp}. Here we continue the discussion by downgrading to the simplest model of a spherical rectangular well, for the first time considered from this perspective in \cite{SchiffSnyderWeinberg1940PhysRev}. We focus on the case of a narrow well, which differs from the modification of the point Coulomb potential at small distances employed in \cite{PomeranchukSmorodinsky1945JPhysUSSR} only by the absence of the Coulomb \enquote{tail}. In addition, in this case, discrete levels are well separated from one another in the entire energy range $-mc^2 \leqslant E \leqslant mc^2$. Therefore, one can easier trace the motion of the poles of the scattering matrix near the boundaries of both the lower and upper continua to make sure that the Breit--Wigner resonances emerging in the vicinity of them are not related to spontaneous electron-positron pair production.
\section{Scattering phases and poles of the $S$-matrix near the boundaries of the upper and lower continua}
\label{sec1}
The radial Dirac equation for states with conserved Dirac quantum number $\kap = \pm 1, \pm 2, \ldots$ in units of $\hbar = c = m = 1$ has the form\footnote{It is identical to the system (1.5.4) in \cite{AkhiezerBerestetskii1981QED} up to the distinction of units after the substitution $rg(r) \rightarrow F(\rho)$, $rf(r) \rightarrow -G(\rho)$ in the latter.}
\begin{equation}\label{eq1}
\begin{gathered}
    H_D \Psi_{\eps, \kap}(\rho) =
        \eps\Psi_{\eps, \kap}(\rho),\quad
    \Psi_{\eps, \kap} = \mqty(F(\rho) \\ G(\rho)),
    \\
    H_D = \mqty(V(\rho) + 1 &
                 \quad \displaystyle{\dv{\rho} - \frac{\kap}{\rho}} \\
                 \displaystyle{-\dv{\rho} - \frac{\kap}{\rho}} &
                 \quad V(\rho) - 1).\quad
\end{gathered}
\end{equation}
Here $\eps$ and $\rho$ are the dimensionless energy and radius; $|\kap| = j + 1/2$, where $j=(l + \wt{l})/2$ is the total angular momentum; $l$ and $\wt{l}$ are the orbital angular momenta of the upper and lower component of the spinor $\Psi_{\eps, \kap}$, respectively: $l = -\kap -1$,\; $\wt{l} = -\kap$ for $\kap < 0$ and $l = \kap$,\; $\wt{l}= \kap - 1$ for $\kap > 0$.

For potential
\begin{equation}\label{eq2}
    V(\rho) = \theta(R - \rho) V,\quad \rho = r / l_C,\quad
        l_C = \hbar/mc,
\end{equation}
where $\theta(x)$ is the Heaviside step function and the depth of the well $V$ is measured in units of $mc^2$, the equation (\ref{eq1}) is solved in terms of Bessel functions \cite{GradshteynRyzhik1980AcadPress_eng}. The partial phase of elastic scattering $\delta_\kap(k)$ is determined by the matching condition
\begin{equation}\label{eq3}
\begin{aligned}
    \frac{J_{j+1}(KR)}{J_{j}(KR)} = &
    \qty[\frac{(\eps + 1)(V + \eps - 1)}{(\eps - 1)(V + \eps + 1)}
        ]^{\frac{1}{2}\sigma}\\
    &\times \frac{J_{j+1}(kR) - \tan\delta_\kap(k)\, N_{j+1}(kR)}
         {J_{j}(kR) - \tan\delta_\kap(k)\, N_{j}(kR)},
\end{aligned}
\end{equation}
where $K = \sqrt{(V + \eps)^2 - 1}$ and $k = \sqrt{\eps^2 - 1}$ are the particle wavevectors inside and outside the well, and $\sigma = \sgn\kap$ is the sign of $\kap$.

As usually, the position of the poles of the partial matrix of elastic scattering $S_\kap(k) = \exp[2i \delta_\kap(k)]$ as a function of the complex variable $k$ is determined by the equation
\begin{subequations}\label{eq4}
\begin{equation}
    \exp[-i\delta_\kap(k)] = 0,
\end{equation}
or
\begin{equation}
    \cot\delta_\kap(k) = i.
\end{equation}
\end{subequations}
In particular, for a discrete spectrum, by specifying $k = i\lambda$, we have \cite{PopovMur1974SovJNuclPhys}
\begin{equation}\label{eq5}
\begin{gathered}
    \frac{J_{j+1}(KR)}{J_{j}(KR)} = -\sigma
    \qty[\frac{(1 + \eps)(V + \eps - 1)}{(1 - \eps)(V + \eps + 1)}
        ]^{\frac{1}{2}\sigma}
    \frac{K_{j+1}(\lambda R)}{K_{j}(\lambda R)},\\
    \lambda = \sqrt{1 - \eps^2},\quad -1 \leqslant \eps \leqslant 1.
\end{gathered}
\end{equation}

A critical value $V_c$ of the well depth, for which a discrete level with $\kap < 0$ crosses the boundary $\eps = -1$, follows from the equation
\begin{equation}\label{eq6}
    J_j(K_c R) = 0,\quad K_c = \sqrt{V_c(V_c - 2)},
\end{equation}
which gives \cite{PopovMur1974SovJNuclPhys}
\begin{equation}\label{eq7}
    V_c(n, \kap < 0) = \sqrt{1 + \qty(\xi_{n,j}/R)^2} + 1,
\end{equation}
where $\xi_{n,j}$ is the $n$-th positive root of the Bessel function $J_j(x)$. At the same time, the depth of the well $V_b$ when the level appears is determined by a more complex equation:
\begin{equation}\label{eq8}
\begin{gathered}
    V_b(n, \kap < 0) = -2 + \frac{(K_b R)}{j}\cdot
        \frac{J_{j+1}(K_b R)}{J_{j}(K_b R)},\\
    K_b = \sqrt{V_b(V_b + 2)}.
\end{gathered}
\end{equation}
On the contrary, for a level with $\kap > 0$, the critical depth of the well is determined by \cite{PopovMur1974SovJNuclPhys}
\begin{equation}\label{eq9}
\begin{gathered}
    V_c(n, \kap > 0) = 2 - \frac{(K_c R)}{j}\cdot
        \frac{J_{j+1}(K_c R)}{J_{j}(K_c R)},\\
    K_c = \sqrt{V_c(V_c - 2)},
\end{gathered}
\end{equation}
while the depth of the well when the level appears reads
\begin{equation}\label{eq10}
    V_b(n, \kap > 0) = \sqrt{1 + \qty(\xi_{n, j}/R)^2} - 1.
\end{equation}

In order to trace the motion of the poles of the scattering matrix near the continua boundaries upon variation of $V$, it is convenient to represent the equation (\ref{eq5}) as an expansion in even and odd powers of $\lambda$ \cite{MurPopov1976TheorMathPhys_FermionCase}. Such an expansion is of the form 
\begin{equation}\label{eq11}
    V - V_b = c_2\lambda^2 + c_4\lambda^4 + \ldots + c_{2l+1}\lambda^{2l+1} +
    \ldots,
\end{equation}
near the boundary of the upper continuum, or
\begin{equation}\label{eq12}
    V_c - V = \wt{c}_2\lambda^2 + \wt{c}_4\lambda^4 + \ldots +
        \wt{c}_{2\wt{l}+1}\lambda^{2\wt{l}+1} + \ldots\, .
\end{equation}
near the boundary of the lower one, respectively.

The condition $k = i\lambda$ fixes the position of the poles of the $S$-matrix in the lower half-plane $k'' > 0$ of the complex variable $k = k' - ik''$, i.e. on the unphysical sheet. These poles correspond to virtual or quasistationary states, which near the continua boundaries manifest themselves as the Breit--Wigner resonances. For potentials $V(r)$, decaying faster than $r^{-2}$ at $r \rightarrow \infty$, the threshold behavior of the widths of such resonances is determined by the permeability of the centrifugal barrier for slow particles,
\begin{equation}\label{eq13}
    D \propto k^{2L+1},\quad k \rightarrow 0.
\end{equation}
Here one should specify an orbital angular momentum $L$ to the orbital angular momentum $l$ of the upper component of the spinor (\ref{eq1}) near the upper continuum boundary, or to the orbital angular momentum $\wt{l}$ of the lower component near the boundary of the lower one, respectively \cite{PopovMur1974SovJNuclPhys}. This explains the particular values of the exponents of $\lambda$ showing up in the leading nonvanishing odd terms in (\ref{eq11}) and (\ref{eq12}).

Let us consider first the simplest cases of $\kap = -1$ and $\kap = 1$, i.e. of $s_{1/2}$- and $p_{1/2}$-states\footnote{Traditionally, classification of states in a spherically symmetric potential is designed according to the values of the angular orbital momentum $l$ of the upper component of the solution (\ref{eq1}) in keeping with the non-relativistic limit.}, which are singled out by the fact that the expansions (\ref{eq11}) and (\ref{eq12}) start with a linear term.
\section{Scattering phases and poles of the partial scattering matrix for $s_{1/2}$-states}
\label{sec2}
With $\kap = -1$ and $j = 1/2$, the equation (\ref{eq3}) is reduced to
\begin{equation}\label{eq14}
    \cot\delta^{(s)}(k) = \frac{1}{kR}\left\{
        1 - \frac{(\eps + 1)}{(V + \eps + 1)}\big[
            1 - KR\cot(KR)\big]\right\},
\end{equation}
where the phase $\delta^{(s)}(k) = \delta_{-1}(k) + kR$, and for the $ns_{1/2}$-levels spectrum according to (\ref{eq5}) we have \cite{AkhiezerBerestetskii1981QED}:
\begin{equation}\label{eq15}
\begin{gathered}
    KR\cot(KR) = -\lambda R - \frac{V}{(1 + \eps)}(1 + \lambda R),\\
    \lambda = \sqrt{1 - \eps^2},\quad -1 \leqslant \eps \leqslant 1.
\end{gathered}
\end{equation}

\begin{figure}
    \centering
    \includegraphics[width=0.8\columnwidth]{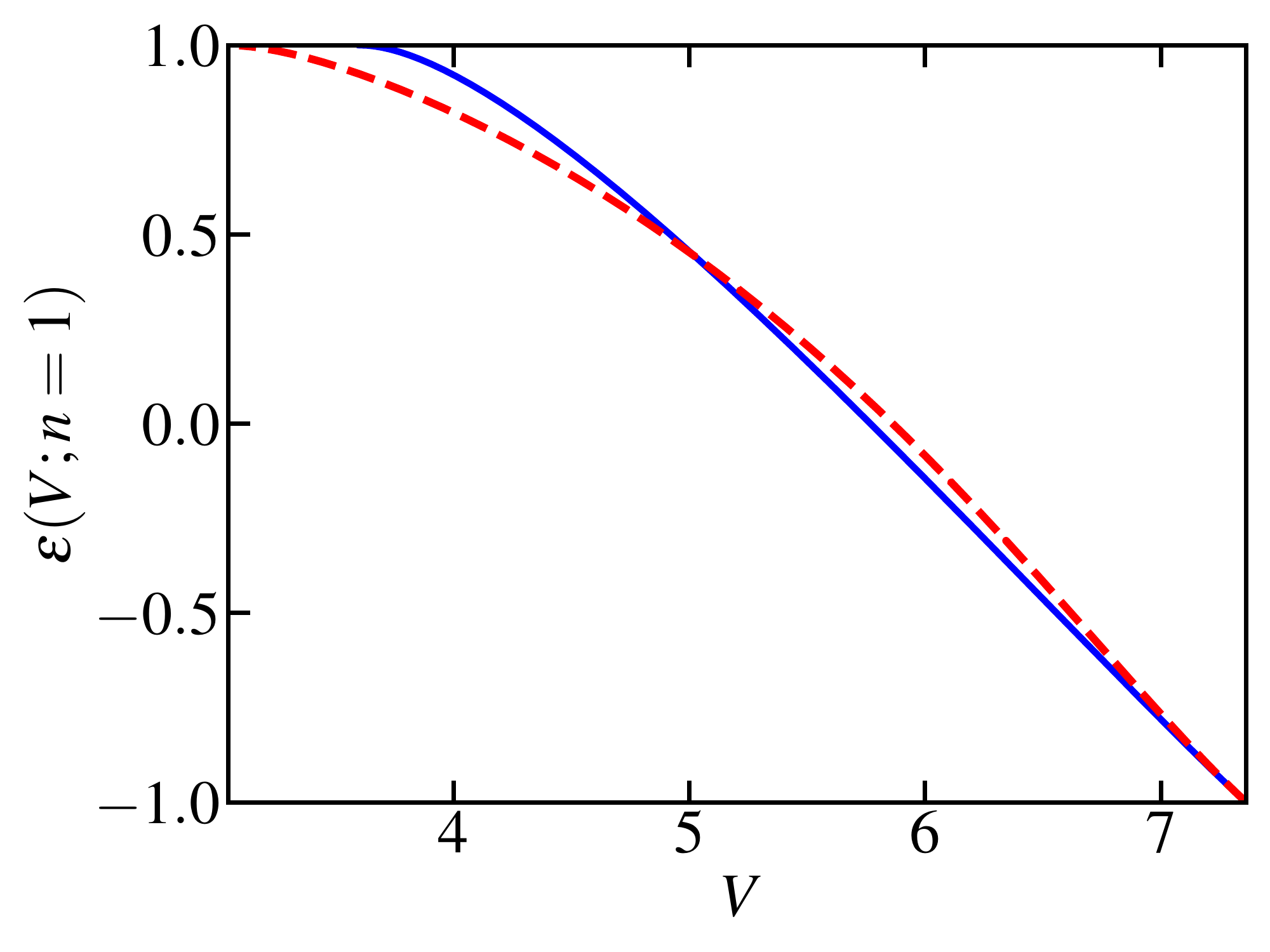}
    \caption{Dependence of the ground ($n = 1$) bound energy level $\varepsilon$ on the depth of the well $V$ at $\kap = -1$ and $R = 1/2$. The solid line corresponds to the numerical solution of the equation (\ref{eq15}), the dashed line~--- to the approximation (\ref{eq16}).}
    \label{fig1}
\end{figure}

For a narrow well, $R \ll 1$, it admits an approximate solution:
\begin{equation}\label{eq16}
\begin{aligned}
    V = &\frac{n\pi}{R} - (2\eps + 1)\\
    &+ \Big[(1 + \eps) \sqrt{1 - \eps^2} +
        \frac{1 - 2\eps(\eps + 1)}{2n\pi}\Big]R+ O\qty(R^2),
\end{aligned}
\end{equation}
where $n = 1, 2, \ldots$ is the radial quantum number. The comparison of the approximation (\ref{eq16}) with the exact (numerical) solution of the equation (\ref{eq15}) is shown in Fig. \ref{fig1}.

\begin{figure}
    \begin{minipage}[b]{1.0\columnwidth}
        \includegraphics[width=1.0\textwidth]{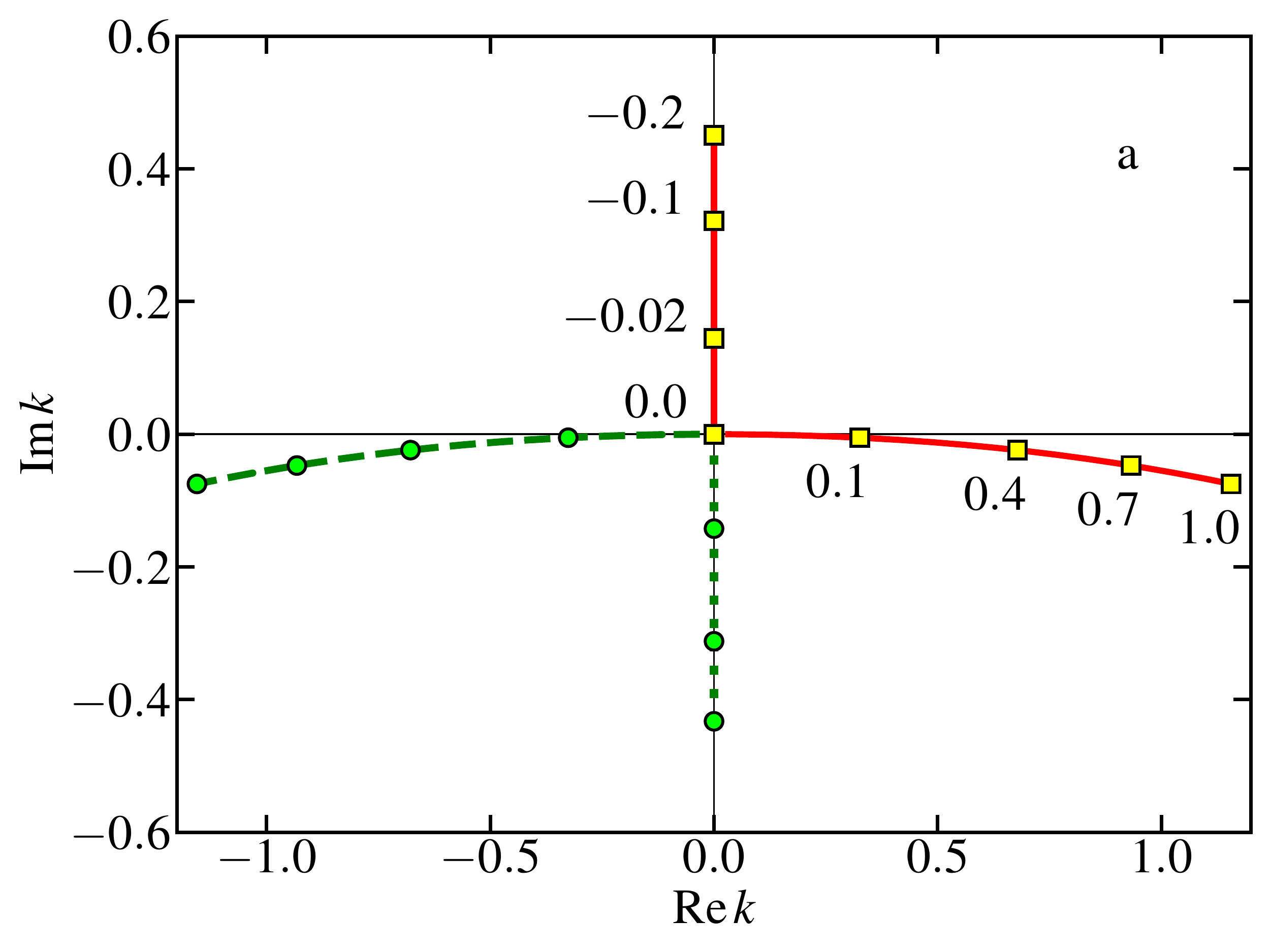}
    \end{minipage}%

    \begin{minipage}[b]{1.0\columnwidth}
        \includegraphics[width=1.0\textwidth]{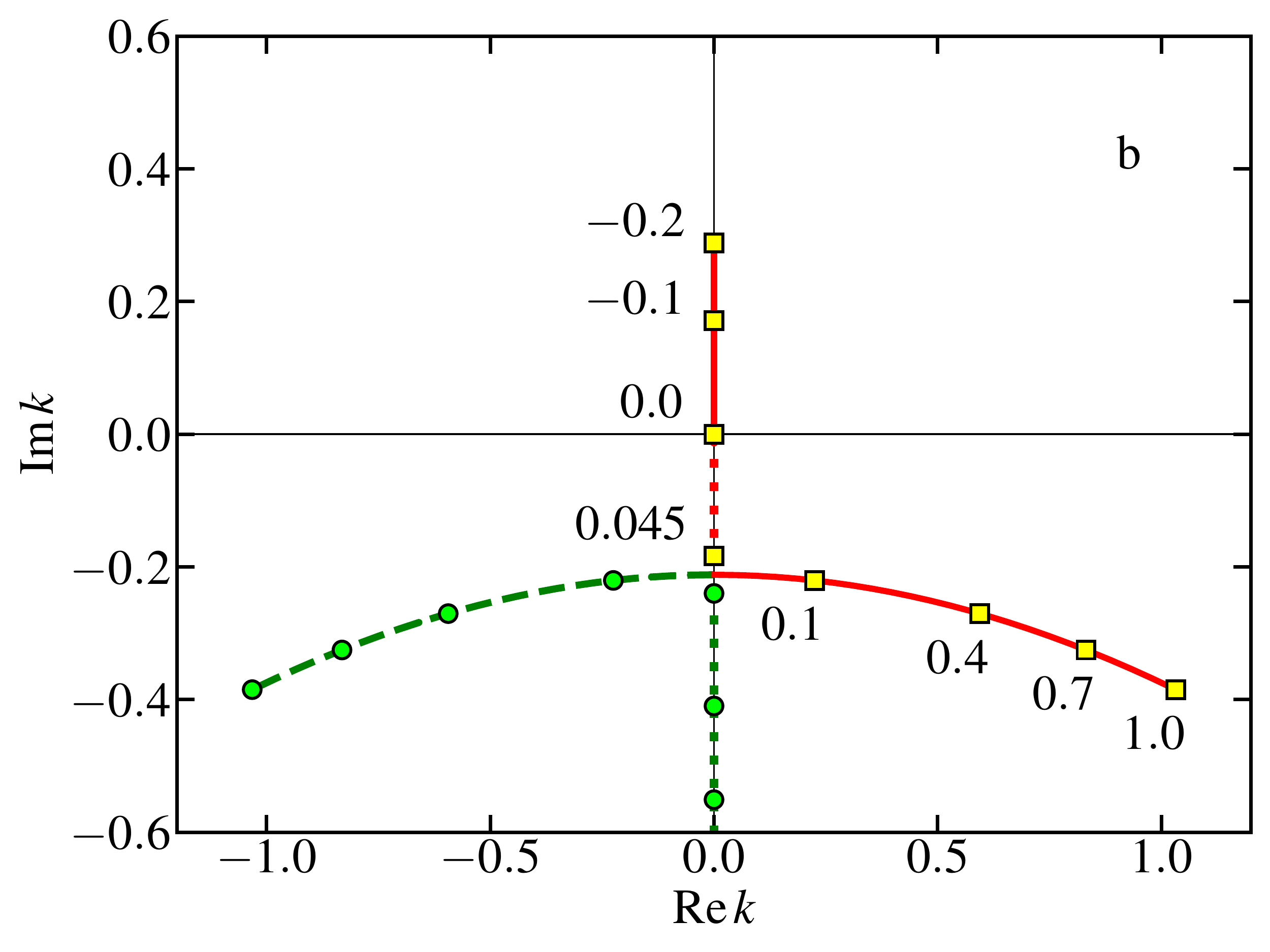}
    \end{minipage}
    \caption{The trajectories of the poles of the $S$-matrix in the complex $k$-plane near the boundaries of the lower~(a) and upper~(b) continua for $\kap = -1$ and $R = 1/5$. The solid line with $\Re k = 0$ corresponds to a discrete level, the dotted line~--- to a virtual one. The solid line with $\Re k > 0$ corresponds to the Breit--Wigner pole, the dashed line with $\Re k < 0$~--- to the second pole, located far from the physical domain. The tick marks indicate the values of $\big(V - V_c^{(s)}\big)$ or $\big(V_b^{(s)} - V\big)$, respectively.}
    \label{fig2}
\end{figure}

By substituting $\eps = -1$ and $n = 1$ into (\ref{eq16}), we arrive at the critical value of the well depth for the ground state,
\begin{equation}\label{eq17}
    V_c^{(s)} = \frac{\pi}{R} + 1 + \frac{1}{2\pi}R + O\qty(R^2),
\end{equation}
which is consistent with the result of \cite{PopovMur1974SovJNuclPhys}. When
$V > V_c^{(s)}$, the ground level \enquote{dives} into the lower continuum, turning into a quasistationary state. To determine the complex energy of such a state, one can use the expansion (\ref{eq12}), which in this case is specified as
\begin{equation}\label{eq18}
\begin{gathered}
    V_c^{(s)} - V = \wt{c}_2^{(s)}\lambda^2 + \wt{c}_3^{(s)}\lambda^3,\\
    \wt{c}_2^{(s)} = 1 - \tfrac{1}{2\pi}R,\quad \wt{c}_3^{(s)} = -\tfrac{1}{2}R.
\end{gathered}
\end{equation}
Analytic continuation to the domain $V > V_c^{(s)}$ results in
\begin{equation}\label{eq19}
\begin{gathered}
    \wt{k}_\pm = \wt{k}'_\pm - i\wt{k}'',\\
    \wt{k}'_\pm = \pm\qty(1 + \tfrac{1}{4\pi}R) \big(V - V_c^{(s)}\big)^{1/2},\\
    \wt{k}'' = \tfrac{1}{4}R \big(V - V_c^{(s)}\big).
\end{gathered}
\end{equation}
The trajectories of the poles of the scattering matrix $S_{-1}(k)$ in the complex $k$-plane parametrized by supercriticality $\big(V - V_c^{(s)}\big)$ are shown in Fig. \ref{fig2}a.

The pole closest to the physical region is called the Breit--Wigner pole, $\wt{k}_\text{BW} = \wt{k}_+$, and the corresponding state in the lower continuum~--- the Breit--Wigner level. For the complex energy of such a level, $\wt{\eps}_\text{BW} = -\sqrt{1 + \wt{k}^2_\text{BW}}$, we have
\begin{equation}\label{eq20}
\begin{gathered}
    \wt{\eps}^{(s)}_\text{BW} = -\wt{\eps}_0^{(s)} +
        \tfrac{i}{2}\wt{\gamma}^{(s)},\\
    \wt{\eps}_0^{(s)} = 1 + \tfrac{1}{2}\qty(1 + \tfrac{1}{2\pi}R)\big(V -
        V_c^{(s)}\big),\\
    \wt{\gamma}^{(s)} = \tfrac{1}{2}R \big(V - V_c^{(s)}\big)^{3/2}.
\end{gathered}
\end{equation}
The unusual sign of the imaginary part, $\wt{\gamma}^{(s)} > 0$, comes due to using a non-second quantized approach to the problem.

\begin{figure}
    \centering
    \includegraphics[width=0.8\columnwidth]{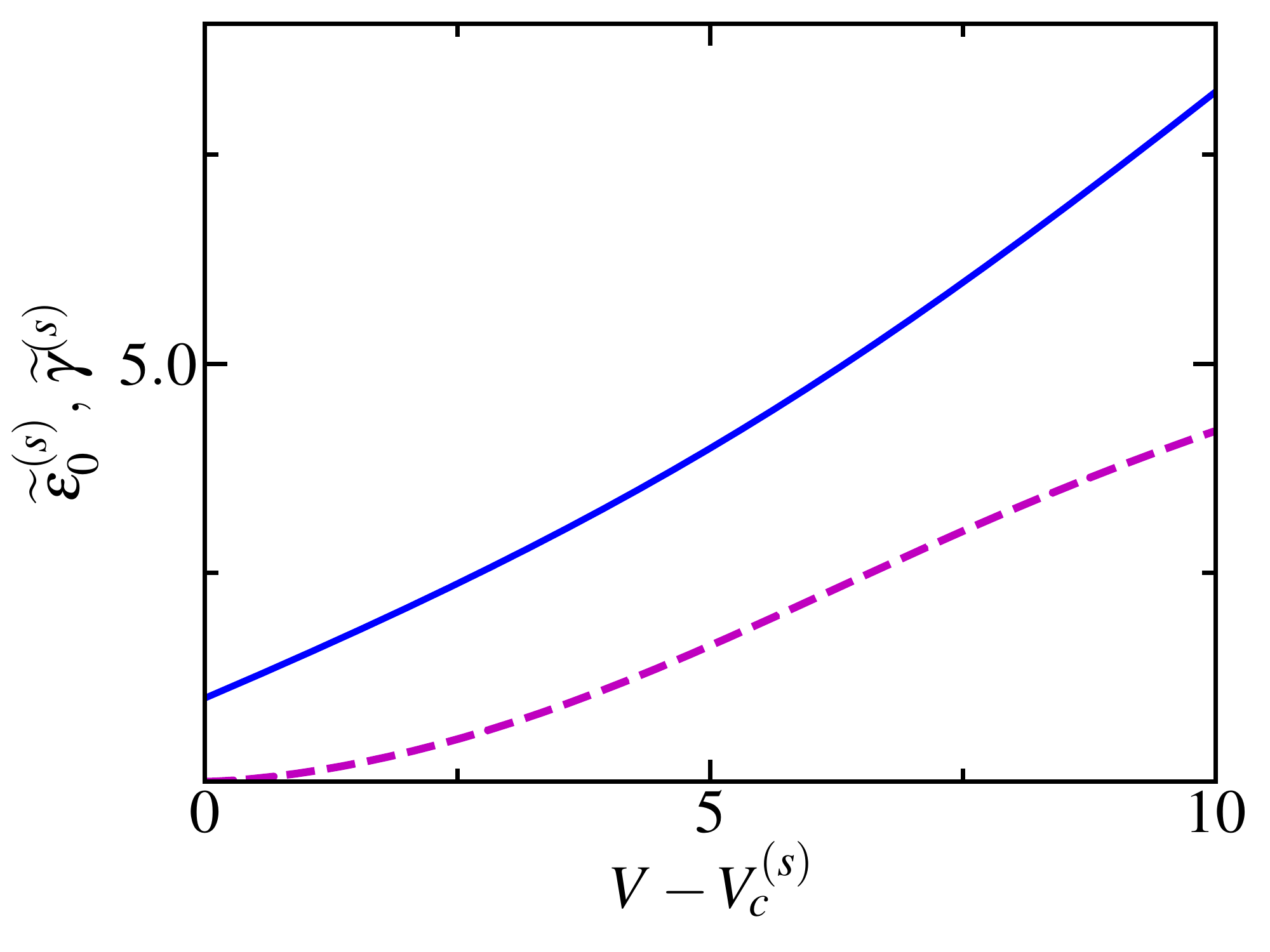}
    \caption{Dependence of the position $\wt{\eps}_0^{(s)}$ (solid line) and the width $\wt{\gamma}^{(s)}$ (dashed line) of the energy level of the positron quasistationary state on supercriticality $\big(V - V_c^{(s)}\big)$ for $\kap = -1$ and $R=1/5$.}
    \label{fig3}
\end{figure}

\begin{figure}
    \centering
    \includegraphics[width=1.0\columnwidth]{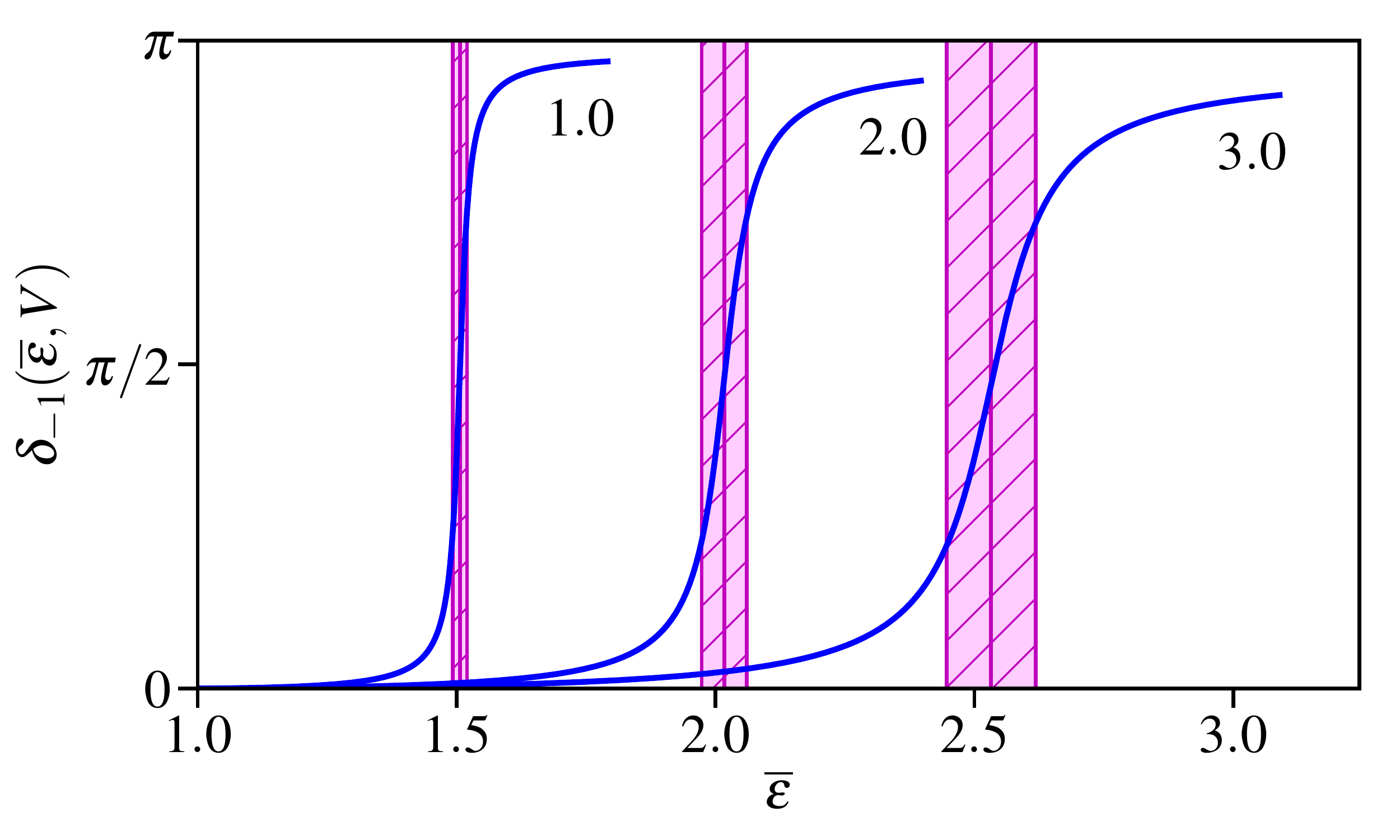}
    \caption{Scattering phases near the boundary of the lower
    continuum for $\kap = -1$ and $R = 1/20$. The shaded areas show the
    positions and widths of the Breit--Wigner resonances. The numbers
    indicate the values of $\big(V - V_c^{(s)}\big)$.}
    \label{fig4}
\end{figure}

A Breit--Wigner level $\wt{\eps}^{(s)}_\text{BW}$ in the Dirac sea corresponds to a quasistationary state of a positron with energy
\begin{equation}\label{eq21}
    \ovl{\eps}^{(s)}_\text{qs} = -\wt{\eps}^{(s)}_\text{BW} =
        \wt{\eps}_0^{(s)} - \tfrac{i}{2}\wt{\gamma}^{(s)},\quad
    \wt{\eps}_0^{(s)} > 0,\quad \wt{\gamma}^{(s)} > 0,
\end{equation}
having the correct sign of the imaginary part. For small supercriticality, this quasistationary state manifests itself as a Breit--Wigner resonance of width $\wt{\gamma}^{(s)}$ in elastic scattering of a positron. The threshold behavior of its width $\wt{\gamma}^{(s)}$ is governed by the permeability of the centrifugal barrier (\ref{eq13}). In the case under consideration, we have $L = \wt{l} = 1$ and $\wt{\gamma}^{(s)} \propto \big(V - V_c^{(s)}\big)^{\wt{l} + 1/2}$, in agreement with the result (\ref{eq20}) given that $k \propto \big(V - V_c^{(s)}\big)^{1/2}$. Figure \ref{fig3} shows the dependence of the level position $\wt{\eps}_0^{(s)}$ and width $\wt{\gamma}^{(s)}$ on supercriticality $\big(V - V_c^{(s)}\big)$, and Fig. \ref{fig4} illustrates the phase of elastic scattering of a positron $\delta^{(s)}(k)$ as a function of its energy $\ovl{\eps} = -\eps$. It is seen that as energy $\ovl{\eps}$ is close to the position of the resonance $\wt{\eps}_0^{(s)}$, the phase changes abruptly across the width $\wt{\gamma}^{(s)}$.

Similarly, one can consider the motion of the poles of the matrix $S_{-1}(k)$ near the boundary of the upper continuum. However, unlike (\ref{eq18}), since the orbital angular momentum of the upper component of the spinor (\ref{eq1}) $l = 0$, here the expansion starts with a linear term 
\begin{equation}\label{eq22}
\begin{gathered}
    V - V_b^{(s)} = c_1^{(s)}\lambda + c_2^{(s)}\lambda^2,\\
    c_1^{(s)} = 2R,\quad c_2^{(s)} = 1 + \tfrac{3}{2\pi}R,
\end{gathered}
\end{equation}
and the first bound level appears at $V = V_b^{(s)} = \pi/R - 3 - 3R/2\pi$, which is consistent with (\ref{eq8}).

Specifying in (\ref{eq22}) $\lambda = -ik$, we obtain
\begin{equation}\label{eq23}
    k_\pm^{(s)} = \pm iR\sqrt{\big(V - V_b^{(s)}\big) \big/ R^2 + 1} - iR.
\end{equation}
As the depth of the well decreases, the discrete level $k_d^{(s)} = k_+^{(s)}$ moves down along the positive imaginary axis of the $k$-plane. When $V = V_b^{(s)}$, it reaches the boundary of the upper continuum, and with a further decrease of $V$ it is pushed into the continuum, turning into a virtual level $k_v^{(s)}$ rather than into a quasistationary state. This is due to the fact that the orbital angular momentum of the upper component of the spinor $l = 0$ and there is no centrifugal barrier.

At the same time, the second virtual level $k_{v'}^{(s)} = k_-^{(s)}$, for which $k_{v'}^{(s)} = -2iR$ at $V = V_b^{(s)}$, moves up towards the level $k_v^{(s)}$ along the same axis. After having collided at the value of the well depth $V = V_b^{(s)} - R^2$, they both go into the complex plane, see Fig. \ref{fig2}b. Note that, surprisingly, the overall pattern of the poles motion near the boundary of the upper continuum for $\kap = -1$ and $R \ll 1$ is quite the same as obtained in  \cite{MigdalPerelomovPopov1971SovJNuclPhys} for nonrelativistic case but with a wide barrier. This apparent puzzle is seemingly explained by the fact that for the discrete spectrum in a narrow well the upper and lower components of the solution (\ref{eq1}) are of the same order of magnitude, so that the situation is far from nonrelativistic one.
\section{Scattering phases and poles of the $S$-matrix for $p_{1/2}$-states}
\label{sec3}
With $\kap = 1$ and $j = 1/2$, the equation (\ref{eq3}) is reduced to
\begin{equation}\label{eq24}
    \cot\delta^{(p)}(k) = \frac{1}{kR}\left\{
        1 - \frac{(\eps - 1)}{(V + \eps - 1)}\big[
            1 - KR\cot(KR)\big]\right\},
\end{equation}
where the phase $\delta^{(p)}(k) = \delta_1(k) + kR$, and for the $np_{1/2}$-levels spectrum according to (\ref{eq5}) we have \cite{PopovMur1974SovJNuclPhys}
\begin{equation}\label{eq25}
\begin{gathered}
    KR\cot(KR) = -\lambda R + \frac{V}{(1 - \eps)}(1 + \lambda R),\\
    \lambda = \sqrt{1 - \eps^2},\quad -1 \leqslant \eps \leqslant 1.
\end{gathered}
\end{equation}
As above, for a narrow well ($R \ll 1$), solution is available via expansion:
\begin{equation}\label{eq26}
\begin{aligned}
    V = &\frac{n\pi}{R} - (2\eps - 1)\\
    &- \Big[(1 - \eps) \sqrt{1 - \eps^2} -
        \frac{1 - 2\eps(\eps - 1)}{2n\pi}\Big]R + O\qty(R^2).
\end{aligned}
\end{equation}

The first bound $p_{1/2}$-level appears when $V = V_b^{(p)}$, where
\begin{equation}\label{eq27}
    V_b^{(p)} = \frac{\pi}{R} - 1 + \frac{1}{2\pi}R + O\qty(R^2),
\end{equation}
and for $V - V_b^{(p)} \ll 1$ equation (\ref{eq25}) is reduced to [cf. (\ref{eq11})],
\begin{equation}\label{eq28}
\begin{gathered}
    V - V_b^{(p)} = c_2^{(p)}\lambda^2 + c_3^{(p)}\lambda^3,\\
    c_2^{(p)} = 1 + \tfrac{1}{2\pi}R,\quad c_3^{(p)} = -\tfrac{1}{2}R.
\end{gathered}
\end{equation}
Comparison with (\ref{eq18}) shows that the motion of the poles of the scattering matrix $S_1(k)$ near the boundary of the upper continuum upon a decrease of $V$ is similar to that of the matrix $S_{-1}(k)$ near the boundary of the lower continuum upon an increase of $V$, see Fig. \ref{fig2}a.

\begin{figure}
    \centering
    \includegraphics[width=0.8\columnwidth]{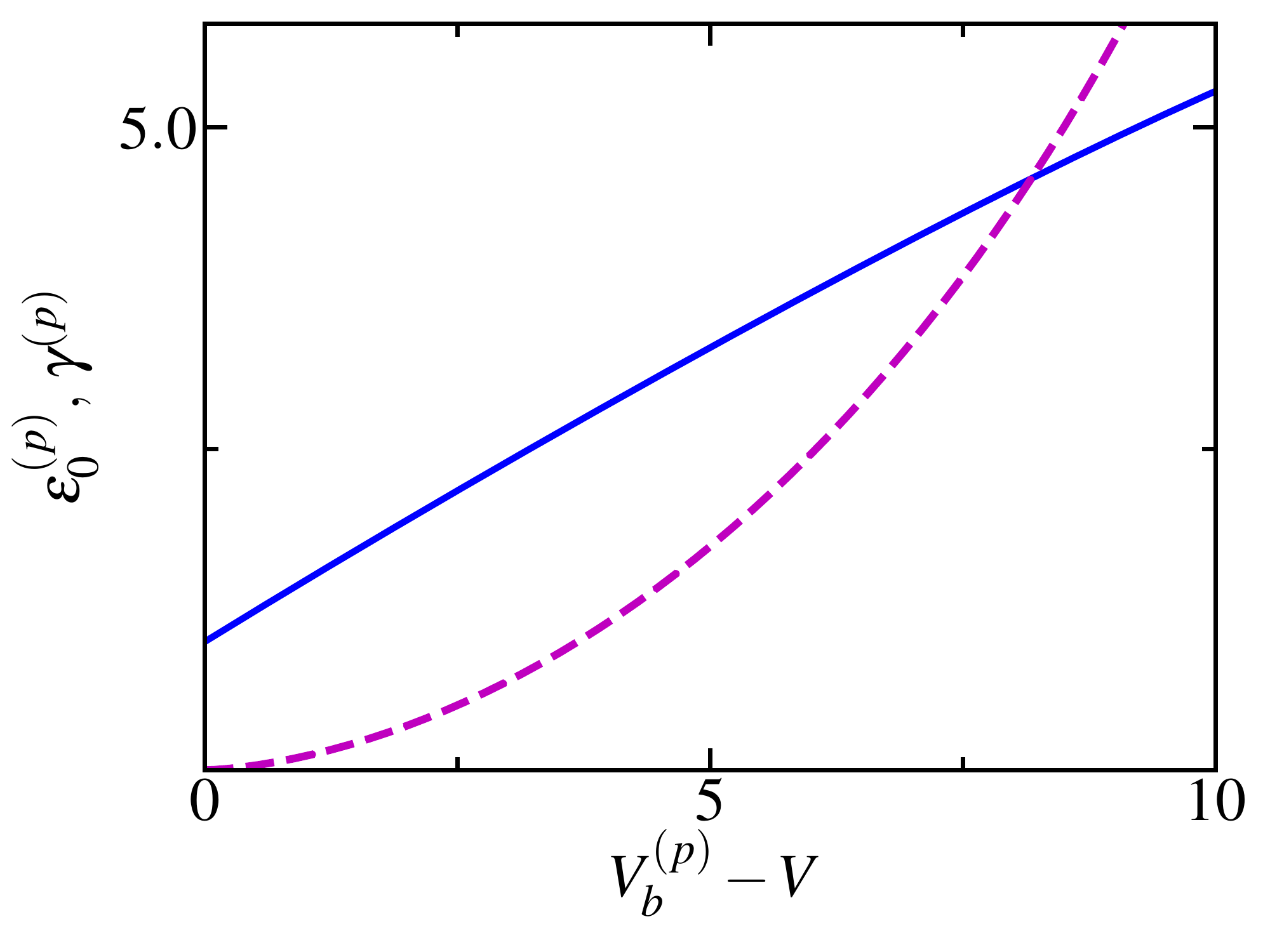}
    \caption{Dependence of the position $\eps_0^{(p)}$ (solid line) and the width $\gamma^{(p)}$ (dashed line) of the energy level of the electron quasistationary state on $\big(V_b^{(p)} - V\big)$ for $\kap = 1$ and $R=1/5$.}
    \label{fig5}
\end{figure}

\begin{figure}
    \centering
    \includegraphics[width=1.0\columnwidth]{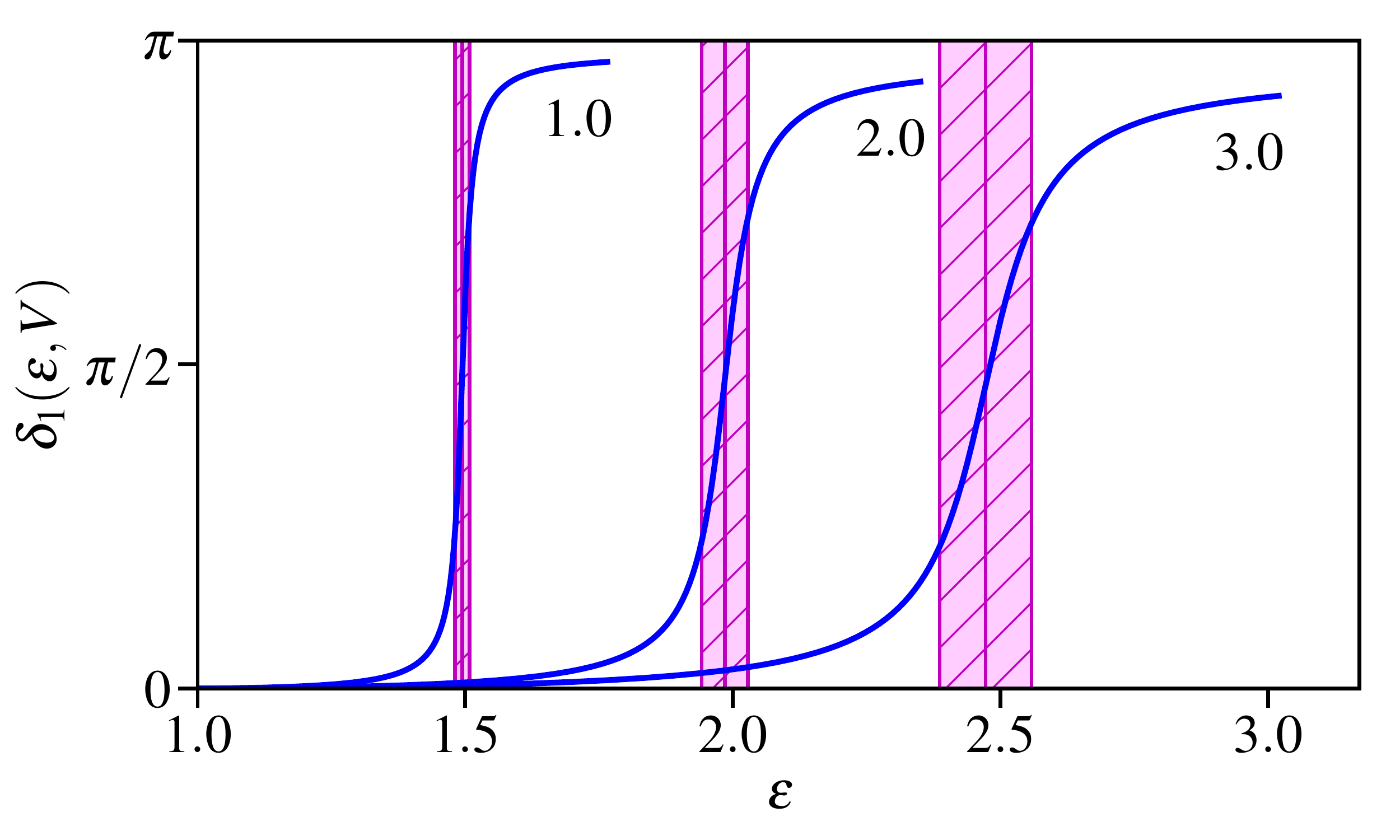}
    \caption{Scattering phases near the boundary of the upper continuum for $\kap = 1$ and $R = 1/20$. The shaded areas show the positions and widths of the Breit--Wigner resonances. The numbers indicate the values of $\big(V_b^{(p)} - V\big)$.}
    \label{fig6}
\end{figure}

For the energy of a quasistationary electron state with $V_b^{(p)} - V \ll 1$, we thus obtain
\begin{equation}\label{eq29}
\begin{gathered}
    \eps_\text{qs}^{(p)} = \eps_0^{(p)} - \tfrac{i}{2}\gamma^{(p)},\\
    \eps_0^{(p)} = 1 + \tfrac{1}{2}\qty(1 - \tfrac{1}{2\pi}R)
        \big(V_b^{(p)} - V\big),\\
    \gamma^{(p)} = \tfrac{1}{2}R\big(V_b^{(p)} - V\big)^{3/2} > 0.
\end{gathered}
\end{equation}
It is worth to emphasize that here, in contrast to the case of levels near the lower continuum boundary, the sign of the imaginary part is usual, i.e. the same as in nonrelativistic theory \cite{Taylor1972}. As previously, the power in the threshold behavior of the width of the Breit--Wigner resonance is ruled by the orbital angular moment $l = 1$ of the upper spinor component, $l + 1/2 = 3/2$. Figure \ref{fig5} shows the position $\eps^{(p)}$ and the width $\gamma^{(p)}$ of the quasistationary state as a function of $\big(V_b^{(p)} - V\big)$, while the phase $\delta^{(p)}(k)$ as a function of electron energy is represented in Fig. \ref{fig6}.

As for the motion of poles with $\kap = 1$ near the lower continuum boundary, it is similar to the one with $\kap = -1$ near the upper continuum boundary. Indeed, in this case $\wt{l} = 0$, hence according to (\ref{eq12}) we have
\begin{equation}\label{eq30}
    V_c^{(p)} - V = \wt{c}_1^{(p)}\lambda + \wt{c}_2^{(p)}\lambda^2,
\end{equation}
and the signs of the coefficients $\wt{c}_1^{(p)}$ and $\wt{c}_2^{(p)}$ are just such that upon $V > V_c^{(p)}$ the $p_{1/2}$-level goes to the second, unphysical sheet.

Indeed, for a narrow well, the equation (\ref{eq26}) gives
\begin{equation}\label{eq31}
\begin{gathered}
    V_c^{(p)} = \frac{\pi}{R} + 3 - \frac{3}{2\pi}R,\\
    \wt{c}_1^{(p)} = 2R,\quad \wt{c}_2^{(p)} = 1 - \tfrac{3}{2\pi}R,\quad R \ll 1,
\end{gathered}
\end{equation}
while for a wide well it gives \cite{PopovMur1974SovJNuclPhys}
\begin{equation}\label{eq32}
\begin{gathered}
    V_c^{(p)} = 2 + \frac{\xi_1^{\;2}}{2R^2},\\
    \wt{c}_1^{(p)} = \frac{\xi_1^{\;2}}{4R^3},\quad
    \wt{c}_2^{(p)} = \frac{1}{2},\quad R \gg 1,
\end{gathered}
\end{equation}
where $\xi_1 = 4,493$ is the least positive root of the equation $\tan\xi = \xi$. Since $\wt{c}_1^{(p)} \ll \wt{c}_2^{(p)}$ for both cases, it is enough to keep just the first two terms in the expansion (\ref{eq12}), which eventually leads  to (\ref{eq30}). Solving for $\lambda$ and specifying $\lambda = -ik$, we obtain
\begin{equation}\label{eq33}
\begin{gathered}
    \wt{k}_\pm^{(p)} = \pm i\sqrt{\nu + \mu^2} - i\mu,\\
    \nu = \big(V_c^{(p)} - V\big) \big/ \wt{c}_2^{(p)},\quad
    \mu = \big(\wt{c}_1^{(p)} \big/ 2\wt{c}_2^{(p)}\big) \ll 1.
\end{gathered}
\end{equation}

As long as $V < V_c^{(p)} + \wt{c}_2^{(p)} \mu^2$, upon the increase of the depth $V$, the real level moves down along the imaginary axis of the $k$-plane, while the second, virtual level moves up towards it. At $V = V_c^{(p)}$ the real level turns into a virtual one passing to the second sheet, but its energy still remains real. When $V = V_c^{(p)} + \wt{c}_2^{(p)} \mu^2$, a collision of the two poles takes place on the second sheet, after which they finally both depart from the imaginary axis. Namely, when $|\nu| \gg \mu^2$, the Breit--Wigner pole with energy
\begin{equation}\label{eq34}
\begin{gathered}
    \wt{\eps}_\text{BW}^{(p)} = -\wt{\eps}_0^{(p)} +
        \tfrac{i}{2}\wt{\gamma}^{(p)},\\
    \wt{\eps}_0^{(p)} = 1 + \tfrac{1}{2}\big(V - V_c^{(p)}\big) \big/ \wt{c}_2^{(p)},\\
    \wt{\gamma}^{(p)} = 2\mu\big[\big(V - V_c^{(p)}\big) \big/ \wt{c}_2^{(p)}\big]^{1/2},
\end{gathered}
\end{equation}
moves to the right, remaining closer to the physical domain, cf. Fig. \ref{fig2}b. This corresponds to the energy of the quasistationary state of a positron with $\kap = 1$
\begin{equation}\label{eq35}
    \ovl{\eps}_\text{qs}^{(p)} = -\wt{\eps}_\text{BW}^{(p)} =
        \wt{\eps}_0^{(p)} - \tfrac{i}{2}\wt{\gamma}^{(p)},\quad
    \wt{\eps}_0^{(p)} > 1,\quad \wt{\gamma}^{(p)} > 0.
\end{equation}
Since the collision of the poles occurs near the physical domain, at a small supercriticality the quasidiscrete state (\ref{eq35}) manifests itself as a Breit--Wigner resonance in scattering of positrons. The width of such a resonance, $\wt{\gamma}^{(p)} \propto \big(V - V_c^{(p)}\big)^{1/2}$, is not related to the centrifugal barrier, which is absent since the orbital angular momentum of the lower component of the spinor $\wt{l} = 0$. However, as already noted above, our relativistic situation is accidentally equivalent to a nonrelativistic one but with a wide barrier, cf. Fig. 2 in the paper \cite{MigdalPerelomovPopov1971SovJNuclPhys} to our Fig. \ref{fig2}b above.
\section{Generalization to the case of arbitrary values $\kap \neq \pm 1$}
\label{sec4}
The simplest cases $\kap = \pm 1$ discussed above are special only by that for $\kap = -1$ the expansion (\ref{eq11}) or for $\kap = 1$ the expansion (\ref{eq12}) starts with a linear term. For $|\kap| \geqslant 2$ both orbital angular momenta $l$ and $\wt{l}$ are nonzero, hence both expansions start with a quadratic term. However, in all other respects the general case looks pretty much the same.

The energies of Breit--Wigner resonances near the boundary of the upper continuum, corresponding to quasistationary states of an electron, in complete analogy with equality (\ref{eq29}) read
\begin{equation}\label{eq36}
    \eps_\text{qs}^{(\kap)} = \eps_\text{BW}^{(\kap)} =  \eps_0^{(\kap)} -
        \tfrac{i}{2}\gamma^{(\kap)},\quad
    \eps_0^{(\kap)} > 1,\quad \gamma^{(\kap)} > 0.
\end{equation}
The threshold dependence of their widths $\gamma^{(\kap)}$ is ruled, as before, by the centrifugal barrier:
\begin{equation}\label{eq37}
    \gamma^{(\kap)} \propto \big(V_b^{(\kap)} - V\big)^{l + 1/2},\quad
    l = j + \tfrac{1}{2}\sgn\kap.
\end{equation}

At the same time, near the boundary of the lower continuum, similarly to (\ref{eq20}), we have
\begin{equation}\label{eq38}
    \wt{\eps}^{(\kap)}_\text{BW} = -\wt{\eps}_0^{(\kap)} +
        \tfrac{i}{2}\wt{\gamma}^{(\kap)},\quad
    \wt{\eps}_0^{(\kap)} > 1,\quad \wt{\gamma}^{(\kap)} > 0
\end{equation}
with the threshold dependence
\begin{equation}\label{eq39}
    \wt{\gamma}^{(\kap)} \propto \big(V - V_c^{(\kap)}\big)^{\wt{l} + 1/2},\quad
    \wt{l} = j - \tfrac{1}{2}\sgn\kap.
\end{equation}
It should be noted that the inequality $\wt{\gamma}^{(\kap)} > 0$ holds for a short-range potential of arbitrary shape. The positivity\footnote{In \cite{ZeldovichPopov1972SovPhysUsp} $\wt{\gamma}^{(\kap)}$ was interpreted as the probability of spontaneous electron-positron pair production.} of $\wt{\gamma}^{(\kap)}$ was established in the paper \cite{MurPopov1976TheorMathPhys_FermionCase} using the effective radius approximation, which was developed therein for the Dirac equation, see also \cite{PopovEletskiiMur1976SovPhysJETP}. By means of this approximation, the expansions (\ref{eq11}) and (\ref{eq12}) were substantiated as well.

An invalid sign in front of $\wt{\gamma}^{(\kap)}$ in (\ref{eq38}) indicates that passage to a second quantized approach is required. Nevertheless, the Dirac radial Hamiltonian (\ref{eq1}) with potential (\ref{eq2}) is a self-adjoint operator. Its eigensolutions form a complete set, which according to Furry \cite{Furry1951PhysRev} can be used to quantize a single-particle system. In the Furry picture, the solutions of the Dirac equation in the lower continuum, in complete analogy with the solutions of a free equation, correspond to the states of a positron with energy $\ovl{\eps} = -\eps > 1$. In a non-second quantized theory, i.e. within the scope of a one-particle approach, they correspond to the states in the \enquote{Dirac sea} distorted by the external field.

Thus, the Breit--Wigner poles with energy (\ref{eq38}) correspond to quasistationary states of a positron with energy
\begin{equation}\label{eq40}
    \ovl{\eps}_\text{qs}^{(\kap)} = -\wt{\eps}_\text{BW}^{(\kap)} =
        \wt{\eps}_0^{(\kap)} - \tfrac{i}{2}\wt{\gamma}^{(\kap)},\quad
    \wt{\eps}_0^{(\kap)} > 1,\quad \wt{\gamma}^{(\kap)} > 0.
\end{equation}
with a negative (as should be) imaginary part. In the case of small supercriticality, such quasidiscrete levels show up as resonances in positron scattering by a supercritical well. If the energy $\ovl{\eps} > 1$ of a positron lies in a region of abrupt change in the partial scattering phase $\delta_\kap(k)$, such as in Fig. \ref{fig4} with $\kap = -1$, then a resonance in its scattering occurs, and the partial cross section is given by the Breit--Wigner formula
\begin{equation}\label{eq41}
    \sigma_\kap(\ovl{\eps}) = \sin^2 \delta_\kap =
    \frac{\big(\wt{\gamma}^{(\kap)} / 2\big)^2}
         {\big(\ovl{\eps} - \wt{\eps}_0^{(\kap)}\big)^2 +
            \big(\wt{\gamma}^{(\kap)} / 2\big)^2}.
\end{equation}

Due to the fact that the partial scattering phases $\delta_\kap(k)$ are real, see (\ref{eq3}) and Fig. \ref{fig4}, the elastic scattering matrix of positrons $S_\kap = \exp[2i\delta_\kap(k)]$ is unitary. Therefore, in accordance with the quantum scattering theory, there are no inelastic processes in the channel with a given $\kap$, including the spontaneous production of electron-positron pairs.

A remarkable property of the Dirac equation should be emphasized. As in the nonrelativistic case, pushing an electron level into the upper continuum upon a decrease of the well depth leads to emergence of a quasistationary state in a scattering of an electron with energy $\eps > 1$. However, diving of the electron level into the lower continuum upon deepening of the well results in emergence of a quasistationary state in a scattering of a positron with energy $\ovl{\eps} = -\eps > 1$.

Therefore, just as resonant scattering of electrons near the threshold of emergence of a bound state, resonant scattering of positrons at small supercriticality cannot serve as evidence in favor of spontaneous production of electron-positron pairs in a supercritical well. A crucial difference between the case of a strong electrostatic field decaying at infinity and a situation arising either in the Klein paradox \cite{Klein1926ZPhys, SauterZPhys} or in a constant uniform electric field, where the spontaneous electron-positron pair production occurs due to the Heisenberg--Euler--Schwinger mechanism \cite{HeisenbergEuler1936ZPhys, Schwinger1951PhysRev}, is that in the latter cases it is never possible to properly separate the upper continuum from the lower one.
\section{Concluding remarks}
\label{conclusions}
As shown, a one-particle approach to the Dirac equation in a narrow spherically symmetric rectangular well remains valid even in the region of a supercritical well depth. Such model reveals a remarkable property of the Dirac equation, which is as follows: while, upon pushing into the upper continuum, a discrete level turns into a quasistationary state in electron scattering, when diving into the lower continuum, it becomes a quasidiscrete level, manifesting itself at small supercriticality as a Breit--Wigner resonance in positron scattering.

This last statement remains valid in a relativistic Coulomb problem with a supercritical nuclear charge $Z > Z_\text{cr}$. The difference between the Dirac equation with the Coulomb potential modified at small distances and the one with a narrow spherical well is that in the former case the smallness of the resonance width is due to the low-permeable Coulomb barrier rather than to the centrifugal barrier \cite{MurPopovVoskresensky1978JETPLett, MurPopov1978SovJNuclPhys}. Therefore, in a weakly supercritical Coulomb problem the Breit--Wigner resonances in positron scattering arise for any value of the Dirac quantum number $\kap$.

In fact, such a view on the phenomena with $Z > Z_\text{cr}$ was confirmed by direct calculations in \cite{GodunovMachetVysotsky2017EurPhysJ}, in particular in Section 3 of that paper, where the \enquote{Dirac equation for positrons} in the Coulomb field of a supercritical nucleus was considered. It was shown there that the Breit--Wigner resonances in positron scattering occur precisely when, upon an increase of the nuclear charge, an electronic level, moving from the boundary of (in this case) lower continuum, reaches the boundary of the upper one, diving into it and turning into a quasistationary state. Nevertheless, as already mentioned, the authors of \cite{GodunovMachetVysotsky2017EurPhysJ} still believe that these resonances are precisely a signature of an electron-positron pair production process, thus sharing the point of view of the authors of \cite{ZeldovichPopov1972SovPhysUsp}.

Since one can hardly hope to study such phenomena experimentally, it makes sense to look for similar phenomena \enquote{in solid-state physics, for example, in the case of dielectrics or semiconductors, the theory\footnote{See, for example, the work of Keldysh \cite{Keldysh1964SovPhysJETP_DeepLevels} on a theory of deep impurity levels in semiconductors, as well as the work \cite{AleksandrovElesin1977SovPhysJETP}, where the impurity states in optical and exciton insulators are considered.} of which has a formal similarity to the Dirac equation}  \cite{PopovMur1974SovJNuclPhys}. Nowadays, numerous theoretical and experimental studies are devoted to a Coulomb problem in graphene, for which the dynamics of charge carriers is described by an effective two-dimensional Dirac equation \cite{CastroNetoGuineaEtAl2009RevModPhys}.

In this case, the Dirac radial Hamiltonian differs from (\ref{eq1}) by replacing\footnote{In a gapless case of massless fermions, a replacement $V(\rho) \pm 1 \rightarrow V(\rho)$ in (\ref{eq1}) should be made.\protect\\ \phantom{In a gapless case of massless fermions, a replacement $V(\rho) \pm 1 \rightarrow V(\rho)$ in (\ref{eq1}) should be made.}} $\kap \rightarrow -J$, where $J = M + 1/2$ is the total angular momentum, and $M = \delta + m$ ($\delta = 0, 1/2$, $m = 0, \pm 1, \pm 2, \ldots$) is the orbital angular momentum, which in the two-dimensional case can be fractional \cite{Pauli1939HelvPhysActa, Winter1968AnnPhys}. A supercritical Coulomb problem in graphene with a gap in the electronic spectrum was considered in \cite{KuleshovMurEtAl2017JETP}, where it was shown that, precisely as in the relativistic case, a one-particle approximation to the effective Dirac equation is still valid for the case of an impurity with a charge $Z > Z_\text{cr}$.

Moreover, it was shown in Refs. \cite{WangWongShytov2013Science, MaoJiangMoldovan2016NaturePhysics} that the positions of resonances calculated for a gapless effective Dirac equation are in agreement with experimental data on the spectra of current-voltage characteristics measured by the scanning tunneling spectroscopy. This confirms the validity of the one-particle approach for the effective two-dimensional gapless Dirac equation with supercritical impurity charge $Z > Z_\text{cr}$, though further discussion of this topic is beyond the scope of this paper.
\begin{acknowledgments}
We are grateful to D.N. Voskresensky, M.I. Vysotsky, S.I. Godunov, B.M. Karnakov, and V.S. Popov for fruitful discussions, and especially to our friend and teacher N.B. Narozhny, who passed away leaving us at an early stage of this work. The work was supported by the MEPhI Academic Excellence Project (Contract No. 02.a03.21.0005) and by the Russian Foundation for Basic Research (Grant No. 19-02-00643a).
\end{acknowledgments}

\end{document}